\documentclass{article}
\usepackage{graphicx}
\begin{document}

\title{Precision Physics at LEP}
\author{Siegfried Bethke\footnote{bethke@mpp.mpg.de}
 \\ Max-Planck-Institute of Physics, Munich, Germany}
\date{{\small to appear in the book:} \\ {\it From my vast repertoire - the legacy of Guido Altarelli}}
\maketitle

\begin{abstract}
As a part of the homage on Guido Altarelli, summarised in the book 
"From my vast repertoire - the legacy of Guido Altarelli"
edited by S. Forte, A. Levy and G. Ridolfi, this contribution
collects some of the technological and scientific highlights 
of precision physics at LEP, the Large Electron-Positron collider operated, 
from 1989 to 2000, at 
the European Laboratory for Particle Physics, CERN.
\end{abstract}

\vskip0.5cm

Scientifically, the notion of $precision$ relates to the reproducibility and
repeatability of a measurement or a prediction.
In practical terms, it relates to the degree of consistency of repeated
measurements, i.e. to the spread of such results around their 
average or mean value.
The term $precision$ $measurements$ usually implies that one
expects significantly $higher$ precision, i.e. smaller spread,
than previous measurements. 
In absolute terms, the attribute $precise$ is often used for
results or predictions with precisions below the per-cent level, but this 
boundary also depends 
on the current and commonly accepted level of consistency.
For instance, measurements or predictions in the field
of Quantum Chromodynamics, the gauge field theory
of the Strong Interaction, are regarded to be $precise$ if
they reach precisions of 1\% or less, while in Quantum Electrodynamics,
"precision" would require levels of at least sub-per-mille to even parts-per-million.

The electron-positron collider LEP, operated at CERN from 1989 to 2000, at 
$e^+e^-$ centre-of-mass energies between 89 and 209 GeV,
fulfilled the definition of "precision" in all technological and scientific 
respects, from the lay-out of the machine, its performance and stability, 
to the detectors, the data analysis and - last not least - to the corresponding 
theoretical calculations, predictions and interpretations which are 
indispensable to extract physical quantities and conclusions from measurements.

Guido Altarelli was a great admirer of the precision reached at LEP,
as is evident from the fact that he himself is author of many summaries and
reviews of "Electroweak Precision Tests" at LEP, see e.g.
\cite{altarelli98,altarelli04,altarelli11}.
Guido also very actively prepared the grounds for performing 
electroweak precision tests at LEP, see e.g. \cite{altarelli82,altarelli89},
and he proposed model-independent methods to detect effects of new physics beyond the
standard model, based on vacuum polarisation effects and precise measurements
at LEP \cite{altarelli91}.

\section{The LEP collider}\label{sec:collider}

\begin{figure}
\centerline{\includegraphics[width=10cm]{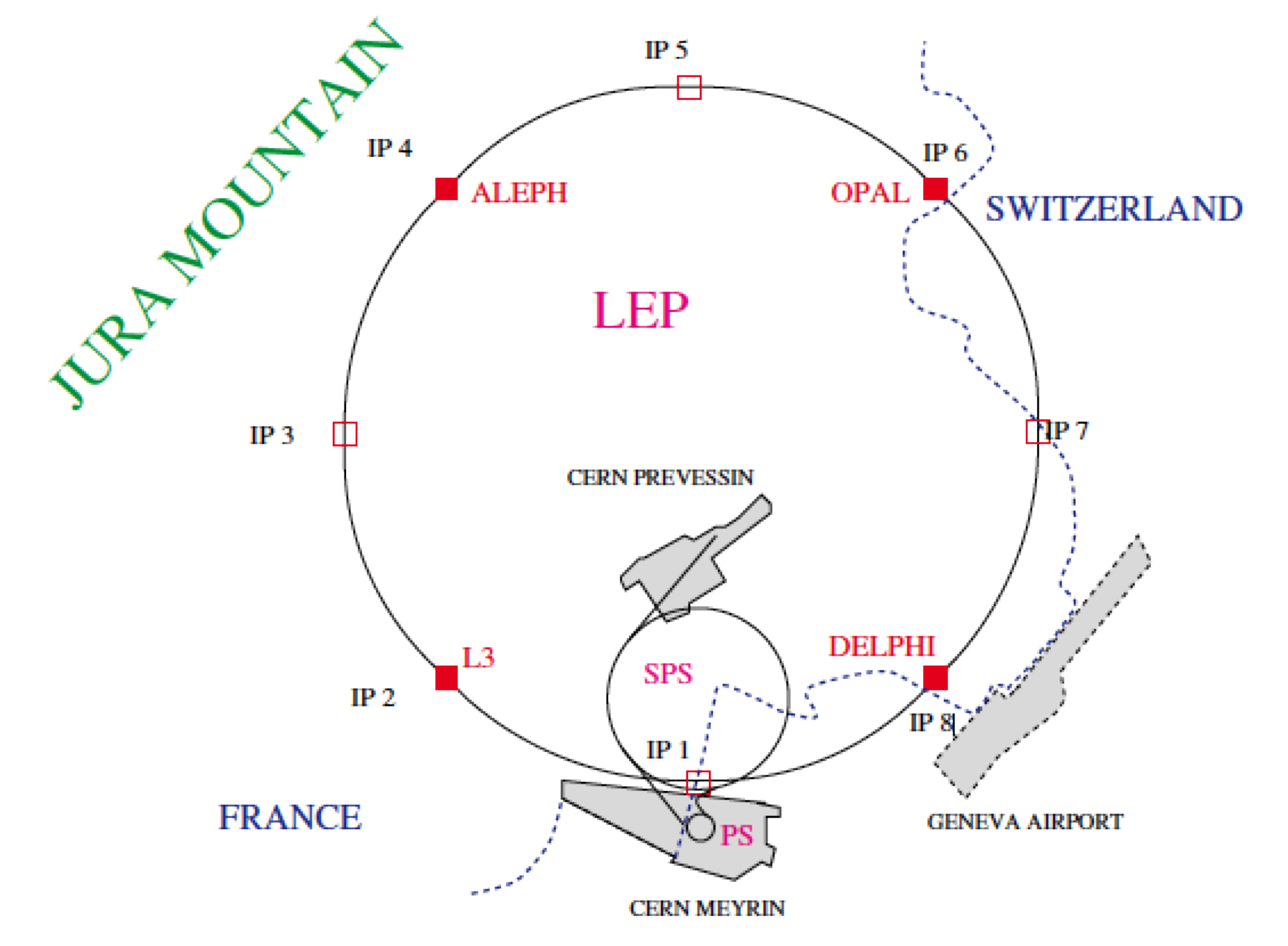}}
\caption{ Layout of the LEP ring and the 8 access points IP1-8,
on the border of France and Switzerland.  
The four LEP experiments L3, ALEPH, OPAL and DELPHI are installed at the
even-numbered access points. Positrons travel clockwise, electrons anticlockwise. 
The location of two LEP injectors, the SPS (Super Proton Synchrotron) and the PS (Protron
Synchrotron), is also indicated. From \cite{brandt00}} \label{fig1}
\end{figure}

The {\bf L}arge {\bf E}lectron {\bf P}ositron Collider was the largest and highest
energy accelerator colliding electrons and positrons ever buildt.
Located in a ring tunnel of 26.67 km circumference about 100m below surface,
LEP operated from 1989 to 2000, providing collision energies from 89 to 209 GeV,
at 4 interaction points equipped with large multi-purpose
particle detectors. 
The layout of LEP is depicted in Fig.~\ref{fig1}.

%
%

Planing, design, approval and construction of LEP commenced from 1975 to 1989,
followed by 11 years of successful operation until LEP's close-down in the fall of 2000.
Initially, 3 phases of operation were planned:  LEP-1 to run at c.m. energies around the
rest-mass of the $Z^0$ boson, $M_Z \approx 91\ $GeV, LEP-2 to run at energies at and
above the $W^+W^-$ production threshold, $2 M_W \approx 160\ $GeV, and finally
LEP-3 as an electron-proton or proton-(anti-)proton collider after installing, additionally,
proton beam acceleration and storage in the LEP tunnel.

\begin{figure}
\centerline{\includegraphics[width=12cm]{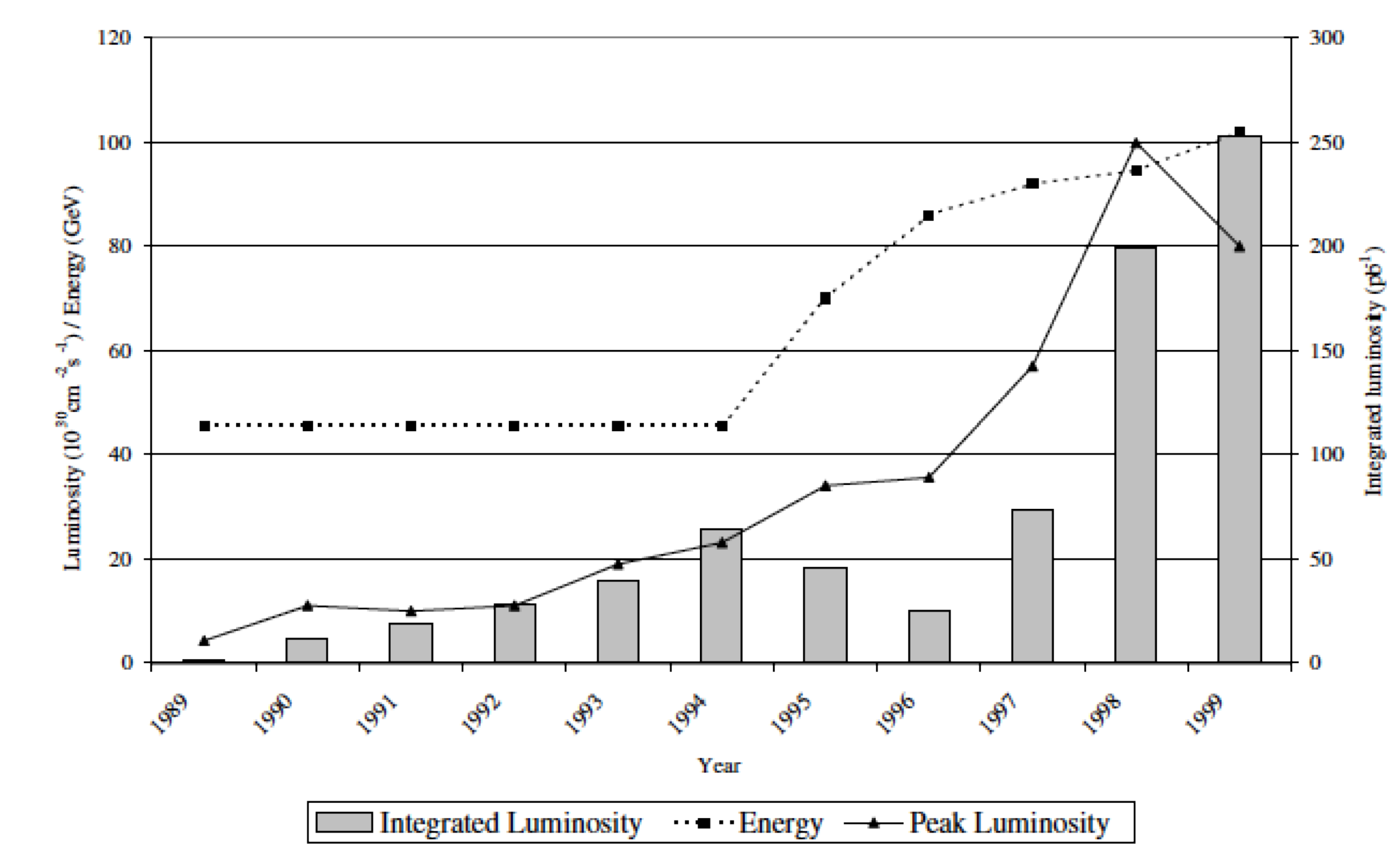}}
\caption{ Maximum energy, maximum luminosity and integrated luminosity per experiment of LEP 
for each year between 1989 and 1999. From \cite{brandt00}} \label{fig2}
\end{figure}

Phase 2 was realised after the development and industrial production 
of superconducting RF cavities, providing nominal acceleration gradients of
6 MeV/m, and their staged installation from 1995 to 1999.
An overview of the maximum energy, maximum luminosities and integrated luminosities 
provided by LEP between 1989 and 1999 is given in Fig.~\ref{fig2}.
Phase 3 was altered and finally realised as the Large Hadron Collider (LHC), after
the shutdown of LEP in late 2000, 
removing the LEP electron-positron infrastructure altogether.
LHC started first operation in 2008, with first proton-proton collisions in 2009.

%
%

The history, design, construction and operation of LEP is documented in a large
number of publications, see e.g. \cite{brandt00,assmann02,huebner04,
schopper09} and references quoted therein. 
Some of the many {\it highlights} of this machine shall be reviewed here.

The most general demands on LEP were to reach and provide highest
energy, luminosity and precision.
The quest for the highest possible beam $energies$ and collision $luminosities$ is 
obvious in terms of determining the parameters of the Standard Model (SM)
of electroweak and strong interactions, i.e. the masses and couplings of gauge bosons (W and Z)
and of the fundamental fermions (quarks and leptons), but also for direct searches of new 
particles and phenomena
(e.g. Supersymmetry) and missing ingredients of the SM (the Higgs boson).
$Precision$ is a key issue for most of these aspects, because precise measurements
of parameters constitute important consistency checks of the underlying theories,
and they provide possibilities to test energy scales much higher than the c.m.
energies of the collisions, through the effects of virtual processes like vacuum polarisation
and vertex corrections.

Among other specifications of the machine, precision knowledge of  the actual beam
energy at the collision points was one of the key issues for LEP operation, most important
for the precise determination of the masses of the Z-boson and the W-bosons.
The initial goal of determining $M_Z$ to 20 MeV (corresponding to
about 0.2 per-mille accuracy) was finally surpassed by a factor of 10, to 20 parts-per-million,
due to the very elaborate and efficient energy calibration and operation of LEP phase-1.

This precision of beam energy calibration was achieved by resonant depolarisation
\cite{assmann99}
of the transverse self-polarisation of the beams that reached, at beam energies around 45 GeV, 
up to 60\%.
This procedure provides accuracies of single measurements to the level of 1~MeV, 
however then unveiled a number of rather subtle and largely unexpected effects
which had to be accounted and corrected for.
Of such, energy deviations due to variations of the circumference of LEP, caused by geological movements due to earth tides and hydrological strains, and effects caused by leakage currents
from the French Train Grande Vitesse (TGV), DC-operated electrical trains circulating 
on nearby railway lines, are perhaps the most remarkable and "entertaining" 
systematic effects, demonstrating the enormous size of precision achieved operating such a large and delicate machine \cite{brandt00}.
These efforts and achievements finally reduced the contribution of the average beam energy 
to the Z-mass uncertainty to 
1.7~MeV (i.e. 18 parts-per-million) and to its overall decay width, $\Gamma_Z$, to 1.3~MeV
\cite{assmann99},

At the higher c.m. energies of LEP phase 2, self-polarisation of the beams could not be established
due to many depolarising effects,
so the calibration of beam energy was performed at lower energies and then transferred "upwards"
using nuclear magnetic resonance (NMR) and flux-loop measurements in the LEP dipole magnets
\cite{assmann05}. 
This provided contributions to the uncertainty of the W mass as low as 10~MeV.

A summary of the most important parameters of LEP operation (phases 1 and 2) is given 
in Table~\ref{tab1}.

\begin{table}[t]
\caption{Basic parameters of LEP}
\begin{center}
{\begin{tabular}{|r|c|c|} 
 \hline
  & LEP-1 & LEP-2 \\ 
 \hline
beam energy up to & 55 GeV & 104.5 GeV \\
 magn.\ field (bending dipoles) & 0.065 T & 0.111 T \\
 accel. voltage per turn & 260 MV & 2700 MV \\ 
 clystron power & 16 MW & 16 MW \\
 RF cavities & Cu (normal condct.) & Cu-Nb (super condct.)\\
  & 128 at P2 and P6 & 272 at P2,4,6,8 \\
  beam currents & 3 mA & 5 mA \\
  number of $e^+e^-$ bunches & 4 x 4 & 4 x 4 (x 2 bunchlets) \\
  max. luminosity & $1.6 \times 10^{31}~cm^{-2}~s^{-1}$ & $5 \times 10^{31}~cm^{-2}~s^{-1}$ \\
  c.m. energy spread & 60 MeV & 140 - 260 MeV \\
  c.m. energy uncertainty & 1.4 MeV & 20-40 MeV \\
  beam lifetime & $\sim 6-8$ h & $\sim 5$ h \\ 
  \hline
\end{tabular}}
\end{center}
\label{tab1}
\end{table}

\section{The LEP detectors}\label{sec:detectors}

Four large general-purpose particle detector systems had been developed, 
approved, constructed and
operated at LEP.
In general, they all shared common features like hermeticity, precision charged particle tracking
and momentum measurement using silicon vertex detectors, gaseous tracking detectors and 
solenoidal magnetic fields, 
particle identification, electromagnetic and hadron calorimeters,
muon tracking and identification, however used different technologies for their subdetectors, magnet systems and detection strategies:
\begin{itemize}
\item 
the ALEPH detector  \cite{aleph90} with it's (at that time) novel and sophisticated
double-sided silicon vertex detector,
\item
the DELPHI detector \cite{delphi91} with emphasis on particle identification by
dE/dx sampling in a large TPC with RICH detectors,
\item
the L3 detector \cite{l390} with a large BGO electromagnetic calorimeter and all
detector components placed inside the world�s largest conventional 0.5T magnet, and
\item
the OPAL detector \cite{opal91} which was based on largely conventional
technologies and anticipated as a safe bet to fully function right from the start of LEP.
\end{itemize}

%
%
\begin{figure}
\centerline{\includegraphics[width=12cm]{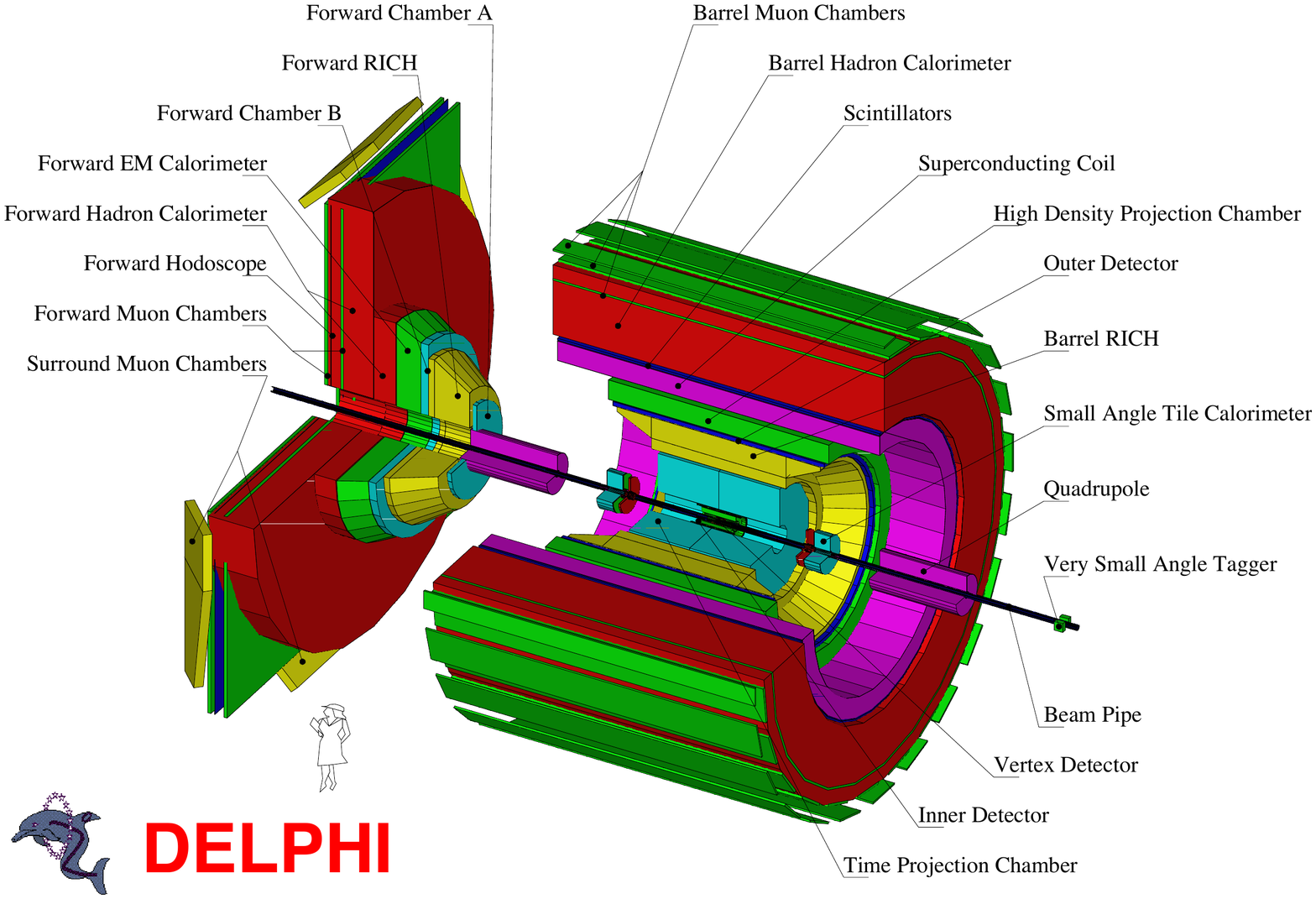}}
\caption{ The DELPHI detector at LEP.} \label{fig3}
\end{figure}

As an example for the size and general structure of the LEP experiments,
a scetch of the DELPHI detector is displayed in Fig.~\ref{fig3}.
A description of all four LEP detector systems would exceed the scope of this
contribution. 
Overviews of the LEP detectors can be found e.g. in \cite{schopper09} and 
\cite{sauli04}. 
Here, and owing to the subject of $precision$ at LEP,
only two of the many innovative technologies and subdetectors shall be
mentioned in more details:
the OPAL Silicon-Tungsten (SiW) luminometer and the ALEPH double-sided
silicon vertex detector.

For precision measurements of production cross sections and (partial) decay widths 
of physical objects like the Z boson, measurements must be compared to well known
"standard" cross sections that are theoretically and experimentally well defined, that can
be measured and monitored with high accuracy, and that are largely independent from any
of the anticipated new physics measurements. At LEP, the measurement of
small angle Bhabha (electron-positron) scattering was chosen to precisely determine 
the luminosity of the LEP beams and 
of recorded data samples, which - at very small scattering angles - 
is basically determined by precision calculations of Quantum Electrodynamics, and is very
little affected by effects of the weak interaction.

%
%
\begin{figure}
\centerline{\includegraphics[width=10cm]{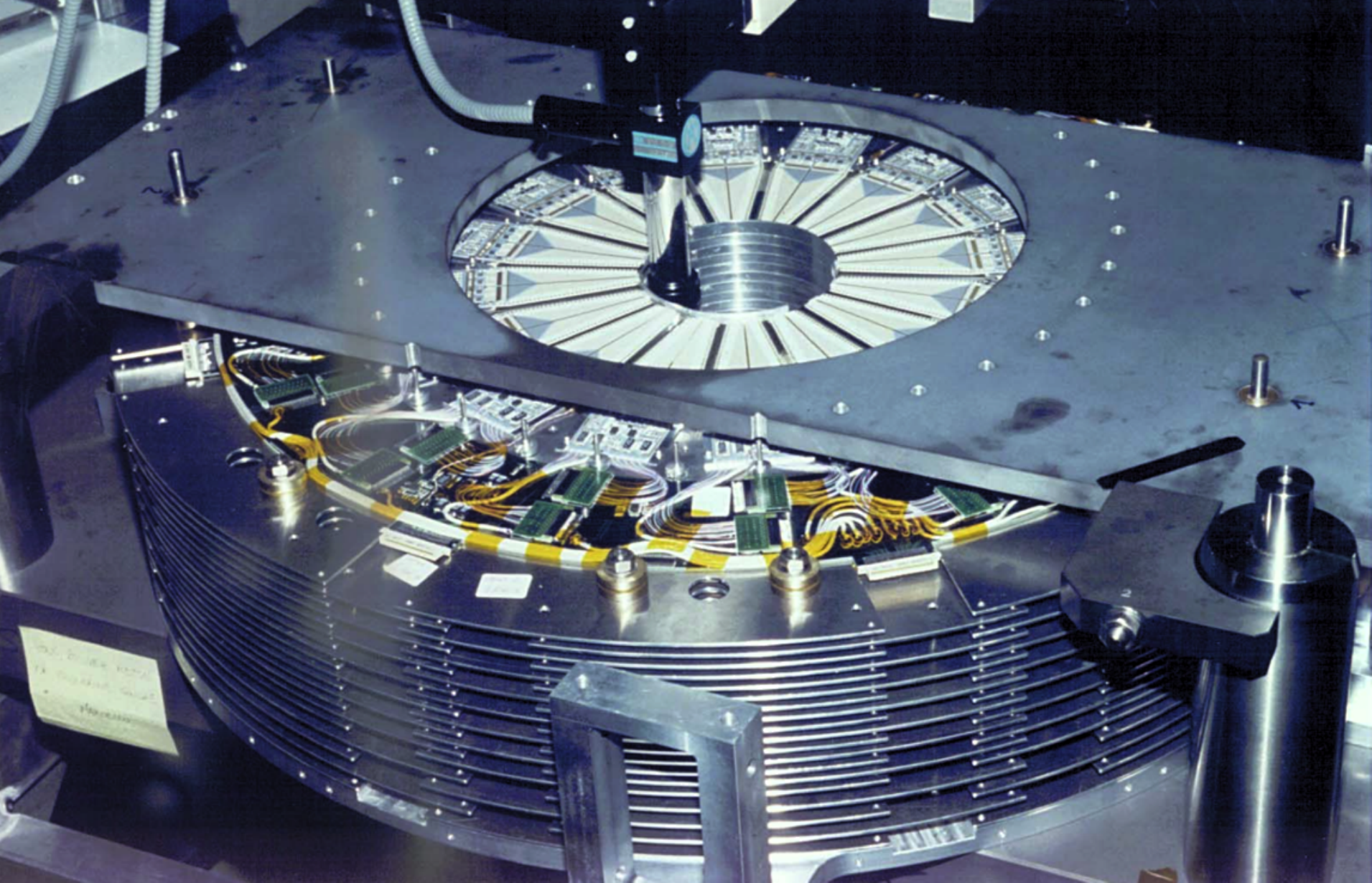}}
\caption{ One of the calorimeters of the OPAL SiW luminometer during construction.
The diameter of the central hole for the beampipe was 10 cm.
The anticipated precision of the luminometer required an accuracy of
the radial positioning of 4.4 $\mu $m.} \label{fig4}
\end{figure}

In the initial phase of LEP-1 running, all 4 experiments had luminometers which determined
beam luminosities to accuracies of about 1\%.
They were soon replaced by more sophisticated and precise devices which achieved
accuracies to about $10^{-4}$.
One of those second-generation luminometers was the silicon-tungsten luminometer
of OPAL \cite{abbiendi99}.
It  was composed of two calorimeters encircling
the LEP beam pipe, on opposite sides of the interaction point,
at approximately �2.5 m from the interaction point,
detecting electrons from
small angle Bhabha scattering at angles between 25 and 58 mrad
w.r.t. the beam pipe.
Each calorimeter was a stack of 19 layers of
silicon sampling wafers interleaved with 18 tungsten plates.
Electromagnetic showers generated in a stack of 1 radiation length of tungsten
absorber plates were sampled by 608 silicon detectors with 38.912 radial pads of 2.5 mm width.
A photograph of one of the calorimeters is represented in Fig.~\ref{fig4}.

The fine segmentation of the detector, combined with the precise 
knowledge of its physical dimensions, allowed
the trajectories of incoming 45 GeV electrons or photons to be determined with a total systematic
error of less than 7 microns.
The total systematic measurement uncertainty was $3.4 \times 10^{-4}$, 
and therefore smaller than the theoretical error of $5.4 \times 10^{-4}$
that was assigned to the QED calculation of the Bhabha acceptance
at that time. 
It thus contributed negligibly to
the total uncertainty in the OPAL measurement of 
the relative invisible decay width of the Z boson, $\Gamma_{inv} / \Gamma_{\ell^+\ell^-}$.

The ALEPH Silicon Vertex Detector \cite{mours96} was the
first detector operating in a colliding
beam environment that used silicon strip detectors which provide readout
on both sides, allowing a three-dimensional point measurement for the
trajectory of charged particles. 
Since then double sided detectors have been applied
in other experiments as well, and have become a standard technology.
Two layers of silicon strip detectors were arranged in two concentric barrels around the beam
pipe with an average radius of 6.3 cm for the inner layer and 10.8 cm for
the outer layer.
The strip detectors had readout strips on both sides. The strips on one side
were parallel to the beam direction while the strips on
the other side were perpendicular to the beam.

%
%
\begin{figure}
\centerline{\includegraphics[width=10cm]{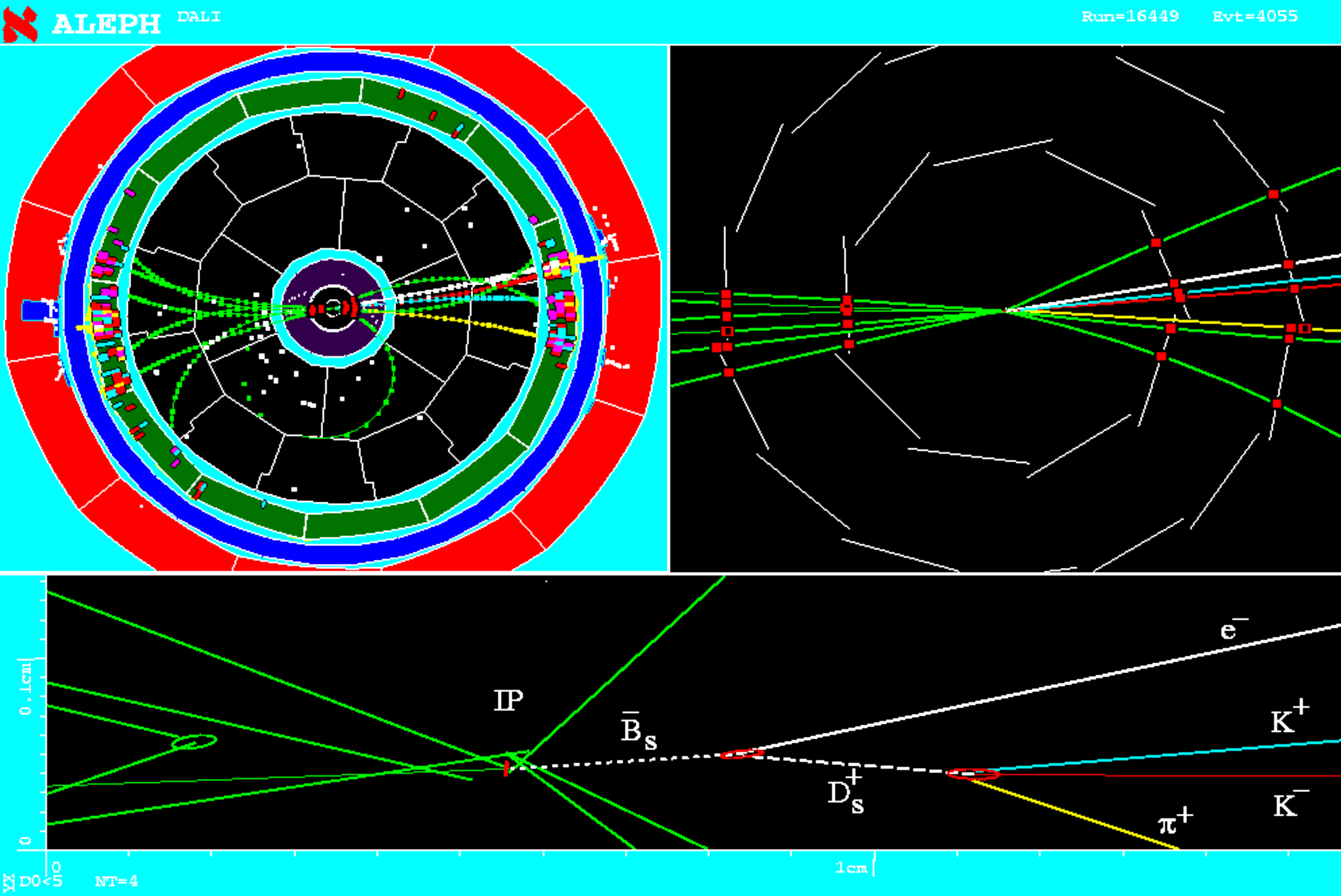}}
\caption{ Candidate for an event $e^+e^- \rightarrow Z^0 \rightarrow b\overline{b}$
as recorded by the ALEPH detector: full detector view perpendicular
to the beam axis (top left), silicon vertex detector with fitted tracks (top right),
and blow-up of reconstructed tracks and decay vertices close to the 
collision vertex (bottom)} \label{fig5}
\end{figure}

The detector system was fully commissioned during 1991 while taking
data at the $Z^0$ resonance. The achieved spatial resolution of the complete
73 728 channel device was 12~$\mu$m in the plane perpendicular to the beam,
and 12~$\mu$m in the direction of the beam. 
This precision, close to the primary collision vertex of LEP,
allowed the measurement of
the lifetimes and decay topologies of short-lived particles
with typical decay lengths down to a few tenths of a millimeter.
An $e^+e^-$ collision event with secondary decay vertices of hadrons containing
b-quarks, recorded by ALEPH, is shown in Fig.~\ref{fig5},

Until 1995, the initial ALEPH silicon vertex detector 
played a vital role in the measurement of $B$-hadron
and $\tau$-lepton lifetimes, of direct $B^0_d$ mixing and limits on $B^0_s$ mixing, 
of $\Gamma_{b\overline{b}}$, and
of limits on the existence of new particles. 
For the further running of ALEPH at LEP phase 2, 
the device was replaced by an advanced version of silicon strip detector,
with larger spacial acceptance, increased radiation hardness and smaller
strip widths.

.
\section{Recap of Standard Model terminology}

The minimal Standard Model (SM) of electroweak interactions
in lowest order ("Born Approximation") describes processes like
$e^+e^- \rightarrow f \overline{f}$ 
using only 3 free parameters:
\begin{itemize}
\item 
$\alpha$ (the electromagnetic coupling $or$ fine structure constant),
\item
$G_F$ (the Fermi constant), and
\item
$\sin^2 \Theta_W$ (the weak mixing angle),
\end{itemize}

\noindent which are fundamental parameters whose numerical values are not given 
by theory but must be determined by experiment.
For instance, $G_F$ and $\sin^2 \Theta_W$ are known
from the muon lifetime and from deep inelastic neutrino-nucleon
scattering experiments, respectively.
Alternatively, $\alpha$, $G_F$ and $M_Z$, the mass of the $Z^0$ boson,
can be chosen as the 3 free and fundamental parameters of the SM, since

$$ 
\sin^2 \Theta_W \ \cos^2 \Theta_W = \frac{\pi \alpha}{G_F \sqrt{2}} \frac{1}{M^2_Z}.
$$

\begin{figure}
\centerline{\includegraphics[width=7cm]{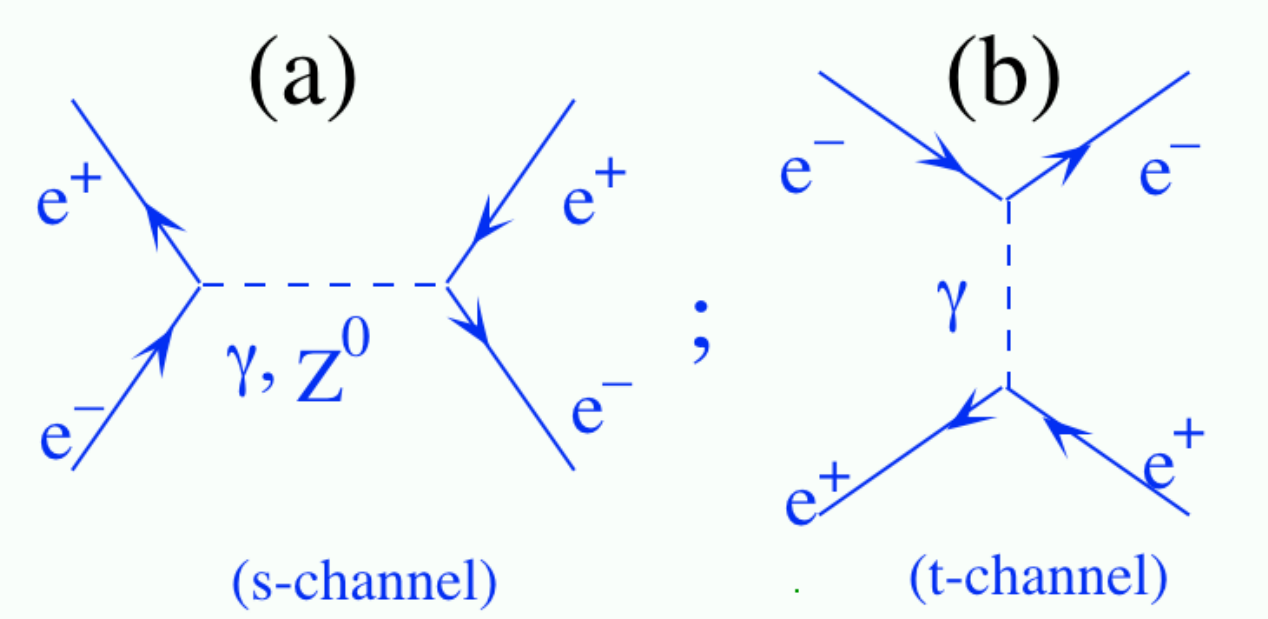}}
\caption{ 
s-channel (a) and t-channel (b) Feynman diagrams for the process $e^+e^- \rightarrow e^+e^-$,
also phrased "Bhabha scattering".
For other charged leptons in the final state, like $(\mu^+ \mu^-)$, $(\tau^+ \tau^-)$ or
(quark-antiquark), only the s-channel annihilation process exists.} \label{fig6}
\end{figure}

The process $e^+e^- \rightarrow f \overline{f}$,
where $f$ can be any of the fundamental fermions of matter, i.e.
$(f \overline{f}) \equiv (e^+e^-),\ (\mu^+ \mu^-),\ (\tau^+ \tau^-),\ 
(\nu_e \overline{\nu_e}),\ ( \nu_{\mu} \overline{\nu_{\mu}}),\ 
(\nu_{\tau} \overline{\nu_{\tau}})$ or 
(quark-antiquark) like $(u \overline{u})$, $(d \overline{d})$, $(s \overline{s}$), 
$(c \overline{c})$, $(b \overline{b})$ and $(t \overline{t})$, is
described by s-channel annihilation diagrams like Fig.~\ref{fig6}a, which proceed via
an intermediate photon ($\gamma$) or a $Z^0$ boson which then branches into the final state 
fermion-antifermion pair.
For the electron-positron final state, also the t-channel scattering diagram given in
Fig.~\ref{fig6}b contributes.
Neutrino final states proceed via s-channel $Z^0$ (all neutrino flavours)
and t-channel $W$ exchange (electron-neutrinos only).

%
%

\begin{figure}
\centerline{\includegraphics[width=9cm]{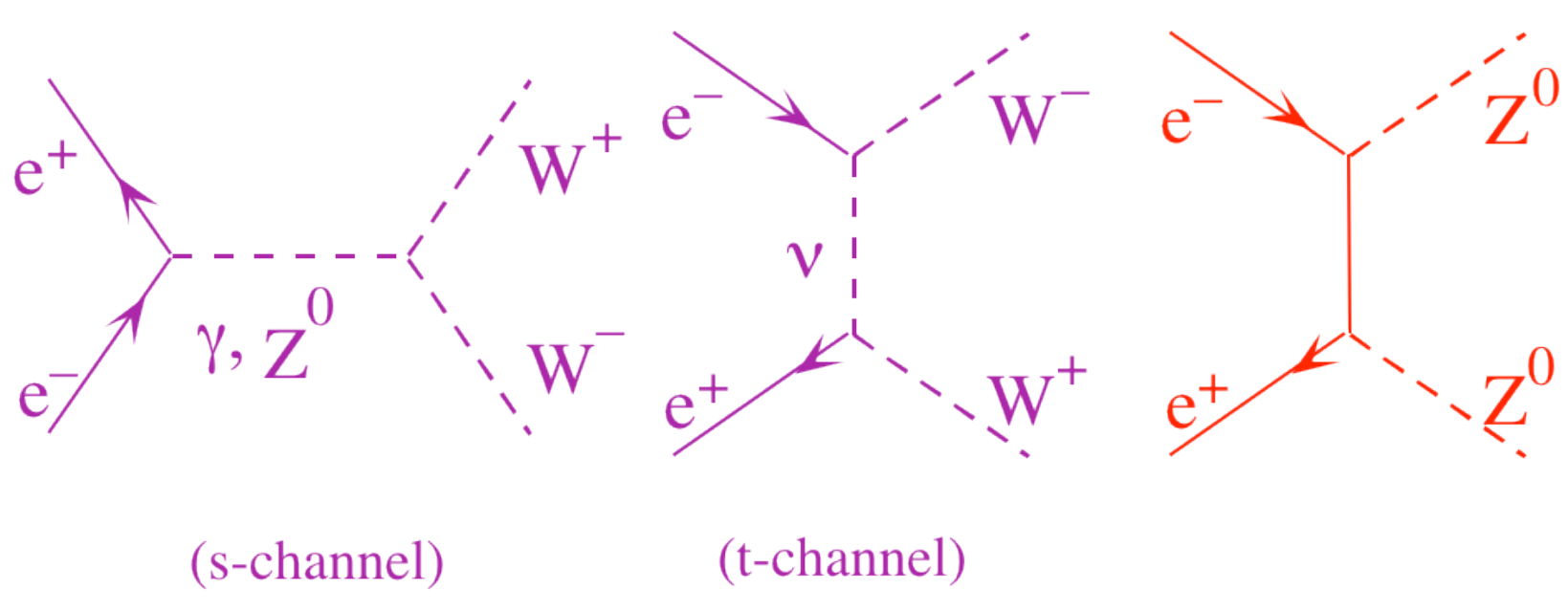}}
\caption{ 
Additional processes and final states in $e^+ e^-$ annihilations
for phase 2 of LEP running, i.e.
at c.m. energies above the $W^+ W^-$ production threshold.} \label{fig7}
\end{figure}

\begin{figure}
\centerline{\includegraphics[width=8cm]{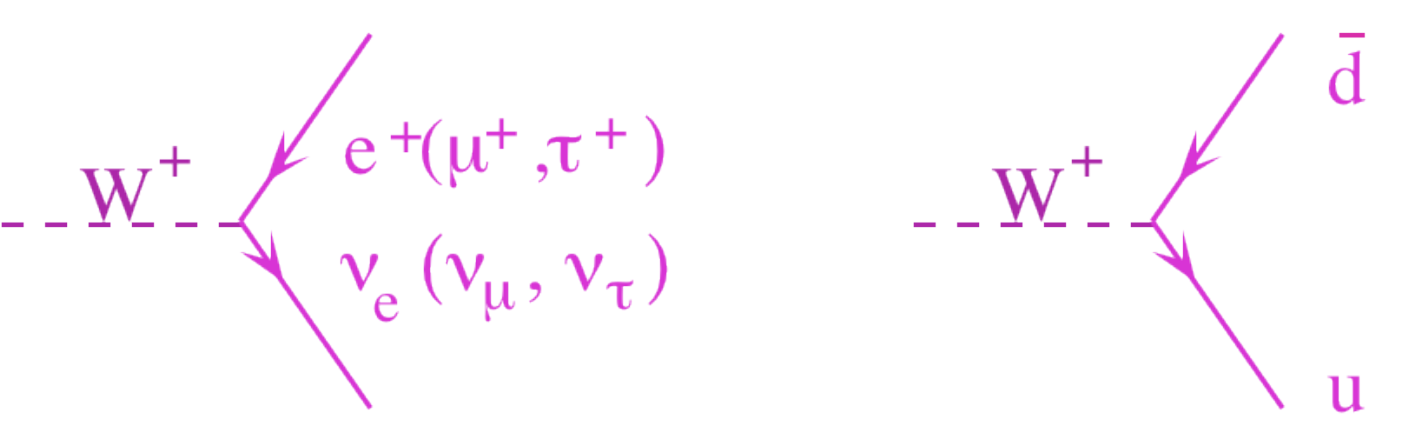}}
\caption{ 
Decay modes of W bosons.} \label{fig8}
\end{figure}

Further diagrams and final states are possible for $e^+ e^-$ annihilation above
the W-boson and Z-boson pair production thresholds, see e.g. Fig.~\ref{fig7}�,
where W bosons instantly  decay into either a lepton and it's associated (anti-)neutrino
or into a quark-antiquark pair as depicted in Fig.~\ref{fig8}, and the Z-boson into a
fermion-antifermion pair (charged leptons, neutrinos or quarks).

%
%
%

%
%

\begin{figure}
\centerline{\includegraphics[width=10cm]{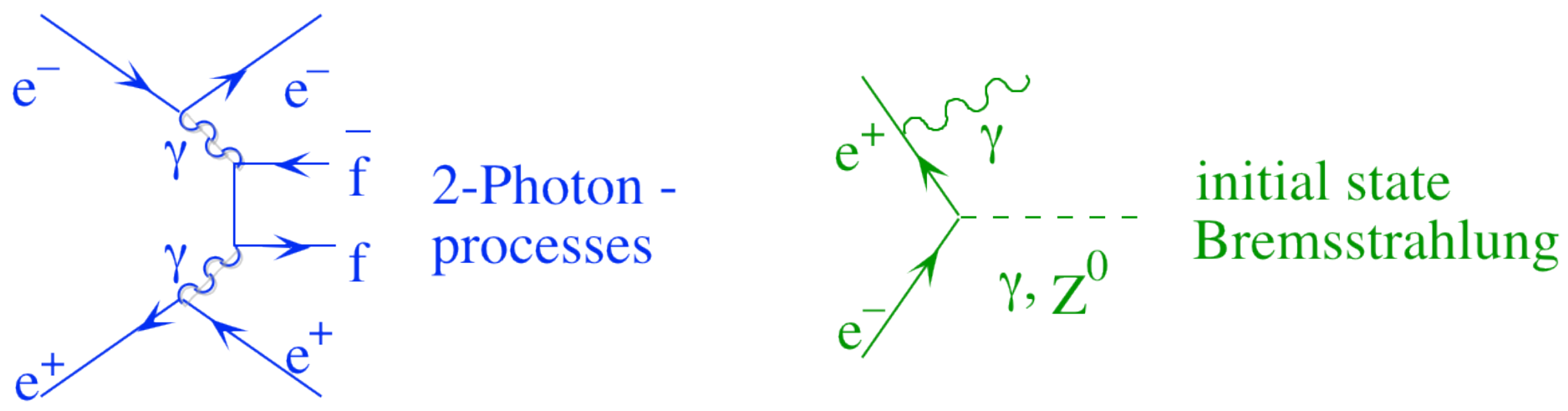}}
\caption{ 
Further processes relevant and produced at LEP.} \label{fig9}
\end{figure}

\begin{figure}
\centerline{\includegraphics[width=10cm]{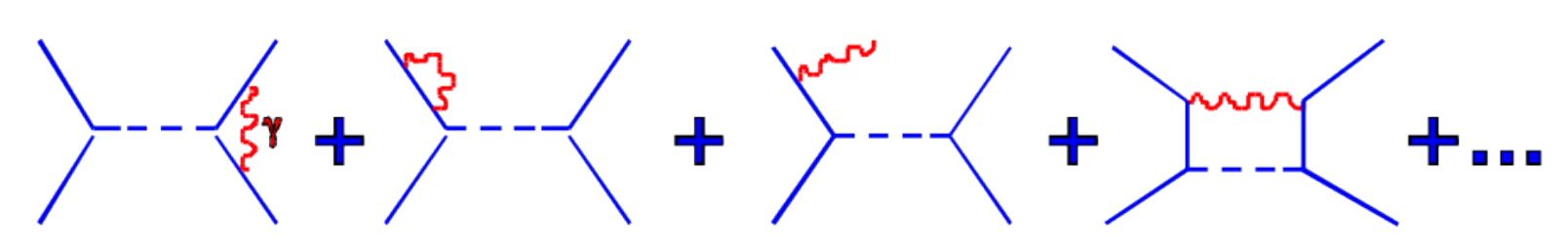}}
\caption{ 
Photonic processes and corrections.} \label{fig10}
\end{figure}

\begin{figure}
\centerline{\includegraphics[width=10cm]{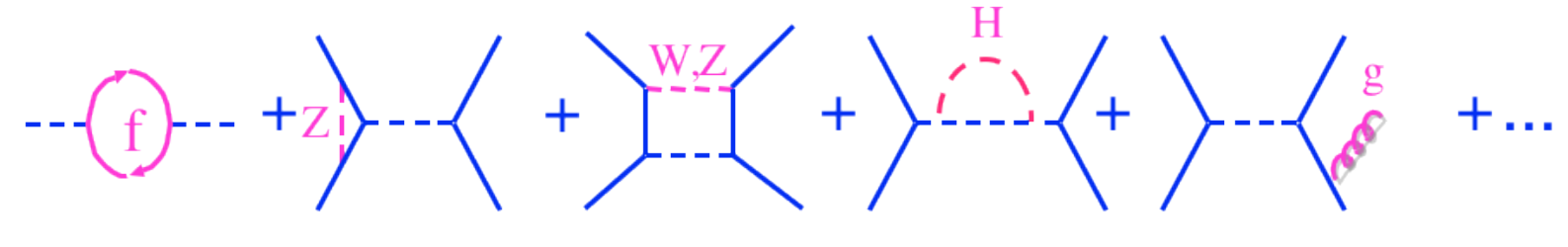}}
\caption{ 
Non-photonic processes and corrections.} \label{fig11}
\end{figure}

Further processes and final states which were relevant and produced at LEP were
e.g. two-photon scattering processes and initial state photon radiation, see
Fig.~\ref{fig9},
and higher-order radiation processes like gluon emission off quarks, final state
photon radiation off charged final state fermions, and virtual processes like 
vacuum polarisation and vertex corrections, some of which are depicted in
Figs.~\ref{fig10} and~\ref{fig11}.

%
%

The production cross sections for fermion-antifermion pairs $f$
in $e^+e^-$ annihilations around c.m. energies of the mass of the
Z boson, 
$\sqrt{s} = E_{cm} \approx M_Z$,  
can be written as
$$\sigma_f(s) = \sigma_f^0 \cdot \frac{s \Gamma_Z}{\left( s - M^2_Z \right)^2
   + M^2_Z \Gamma_Z^2} + "\gamma" + "\gamma Z" \ ,$$
 where the first term parametrises the resonant cross section around the Z mass,
 the second term, $"\gamma "$ is the electromagnetic cross section for point-like 
 annihilation processes - a function falling like $1 / s$ - and $"\gamma Z"$ denotes 
 electro-weak interference terms.
 $\sigma^0_f $ are the pole cross sections for fermions $f$,
 $$ \sigma^0_f = \frac{12 \pi}{M^2_Z} \frac{\Gamma_e \Gamma_f}{\Gamma^2_Z\ ,}$$
 and $\Gamma_Z$, $\Gamma_f$ and $\Gamma_e$ are the total width and the partial
 widths of the Z into fermions $f$ and electrons, respectively.
 
 Measurements of s-dependent cross sections around the resonance of the Z-boson 
 provide $model$ $independent$ results for $M_Z$, $\Gamma_Z$, $\Gamma_f$ and
 $\sigma_f^0$.
 In the SM, $\Gamma_f$ are no free parameters, but are parametrised as functions
 of the so-called vector and axial-vector constants $g_{a,f}$ and $g_{v, f}$:
 $$
 \Gamma_f = \frac{G_f M^3_Z}{6 \pi \sqrt{2}} \left[ g_{a,f}^2 + g_{v,f}^2 \right] N_{c,f}
 $$
where $N_{c,f} = 3$ for quarks and 1 for all other
fermions;  $g_{a,f} = I_{3,f}$ and $g_{v,f} = I_{3,f} - 2 Q \sin^2 \Theta_W$
with $I_{3,f}$ the $3^{rd}$ component of the weak isospin ($= \pm 1/2$) and
$Q$ the electric charge of the fermion.

Radiation corrections in the SM are treated in two different classes:
\begin{itemize}
\item
photonic corrections, cf. Fig.~\ref{fig10}, may be large and cause corrections up to 
${\cal O}(100\%)$; they may depend on selection criteria and their effects can thus be
reduced; and they are factorisable, i.e. their contributions can be factored out 
by expressions like $(1 + \delta_{rad})$.
\item
nonphotonic correctiosn, c.f. Fig.~\ref{fig11}, involving loops, vertex corrections etc. from gluons,
fermions, W, Z and Higgs bosons, even if they are too heavy to be produced as "free" particles
at the given collider energies.
These corrections are in general smaller than the photonic ones, of ${\cal O}(10\%)$, but they are independent of selection criteria. However, they can be absorbed in running couplings.
\end{itemize}

%
%
%

%
%

Higher order radiation corrections render couplings and parameters of the SM to
become energy dependent, and the common notion is to deal with
an effective weak mixing angle,
$\sin^2\theta_{eff}(s)$,
with the electromagnetic coupling $\alpha (s) = \alpha / (1 - \Delta \alpha )$ where 
$\Delta \alpha = 0.064$ at $\sqrt{s} = M_Z$, 
with $N_{c,f} \left(1 + \frac{\alpha_s}{\pi} + 1.4 \left( \frac{\alpha_s}{\pi}\right)^2 + ...\right)$
for quarks, modified by the running strong coupling constant, $\alpha_s (s)$,
and  with $M^2_W / M^2_Z = \rho \cos^2 \theta_w$ with $\rho = 1 / (1 - \Delta \rho)$,
whereby 
$$
\Delta \rho = 0.0026 \frac{M^2_t}{M^2_Z} - 0.0015 \ln \left( \frac{M_H}{M_W}\right).
$$

Inserting the running couplings into the "Born" approximation, the partial Z decay widths
$\Gamma_f$, the cross sections and other observables acquire dependences on 
\begin{itemize}
\item
the top quark mass $M_t$,
\item
the Higgs boson mass $M_H$, and
\item
the strong coupling parameter $\alpha_s$.
\end{itemize}

This then provides the possibility for indirect determinations (i.e. fits) of $M_t$, $M_H$ and $\alpha_s$
from a combination of measurements of electroweak observables.
From the numerical structure of the $\Delta \rho$ parameter given above it is quite obvious
that such determinations, for them to provide significant results, must be based on
$very$ $precise$ measurements: taking the top-quark and Higgs-boson masses as we know
them today, from direct production and measurements of these objects, $M_t = 173.1 \pm 0.6$~GeV
and $M_H = 125.09 \pm 0.24$~GeV \cite{patrignani17}, leads to contributions to $\Delta \rho$
of 0.009 and -0.0012, respectively.
Aiming, for example, at relative uncertainties of $5 \%$ on $M_t$ and $10 \%$ on $M_H$,
in indirect determinations from fits to precision electroweak data alone, requires
control of all statistical and systematic errors to levels corresponding to 
$10^{-3}$ and $5 \times 10^{-4}$, respectively\footnote{These numbers are given for illustration
only.}.

More detailed account of the basic theory, 
calculations of higher order QED, electroweak and
QCD (strong interaction) corrections, Implementations into Monte-Carlo programs, fitting
algorithms and analysis procedures is given in the final reports of the LEP Electroweak Working Group
\cite{schael05,schael13} and respective web updates
\cite{lepewwg12}, as well as in many summaries and reviews as those given by Guido
himself, e.g.  \cite{altarelli98,altarelli04}, and references quoted therein.

\section{Precision electroweak results of LEP}

One of the main tasks of LEP phase 1 experimental program was the 
measurement and determination of the essential parameters
of the Z resonance, its mass, its width, its branching fractions, and the angular distribution
of its decay products.
About 17 million Z decays were accumulated by the four experiments at LEP.
In most of the combined and global analyses of these data, LEP data were 
supplemented by the results from the SLD detector at the Stanford 
Linear Accelerator (SLC).
SLD collected about 600.000 Z decays overall, however with electron beam polarisation
of up to about 80\%, adding additional means for electroweak precision tests that
LEP could not provide.

The final results of electroweak precision measurements around the Z pole
were presented in \cite{schael05}.
Owing to the precision of the LEP beam energies, the mass and total width of
the Z boson were determined with unprecedented precision:

\begin{eqnarray}
M_Z &=& 91.1875 \pm 0.0021\ {\rm GeV}   \nonumber \\
\Gamma_Z &=& 2.4952 \pm 0.0023\ {\rm GeV}, \nonumber
\end{eqnarray}

\noindent
corresponding to a precision of 23 parts-per-million for $M_Z$ and 0.9 per-mille
for $\Gamma_Z$.

\begin{figure}
\centerline{\includegraphics[width=8.3cm]{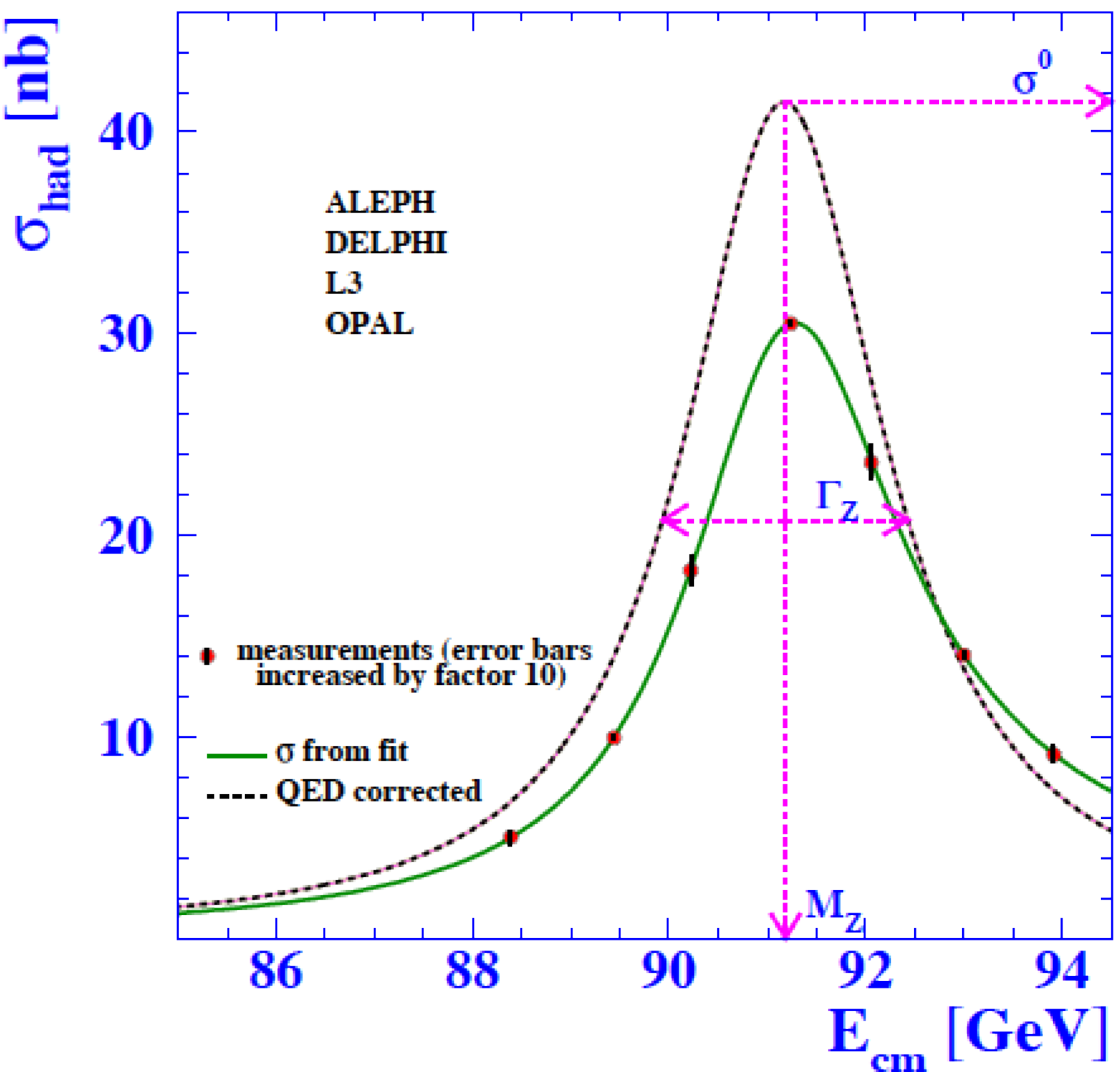}}
\caption{ 
Average of measurements of the hadronic cross-section as a function of 
$e^+e^-$ centre of mass energy.
The full line is a model independent fit to the measurements; the dashed
line is obtained by correcting the former for QED photonic effects. 
From \cite{schael05} } \label{fig12}
\end{figure}

One of the key measurements, the cross section for $Z^0 \rightarrow$\ 
hadrons\footnote{About 70\% of all Z bosons decay hadronically, into a
quark-antiquark pair.}, is given in Fig.~\ref{fig12}, form which the basic parameters
of the Z-boson are being derived.
A particular important derivation of these data, with global and even cosmological relevance, 
is the determination of the number of light neutrino flavours coupling to
the Z-boson, $N_\nu$.
It is derived from the Z invisible partial decay width, $\Gamma^{inv}_Z$,
which is determined from the measurements
of visible Z decays into charged leptons and quarks, and results in
$$
N_\nu = 2.9841 \pm 0.0083,
$$
\noindent
assuming lepton universality and excluding other contributions to $\Gamma^{inv}_Z$
than SM neutrinos.
It is therefore concluded that only the three known sequential generations of fermions
exist in nature.

%
%

\begin{figure}
\centerline{\includegraphics[width=6.4cm]{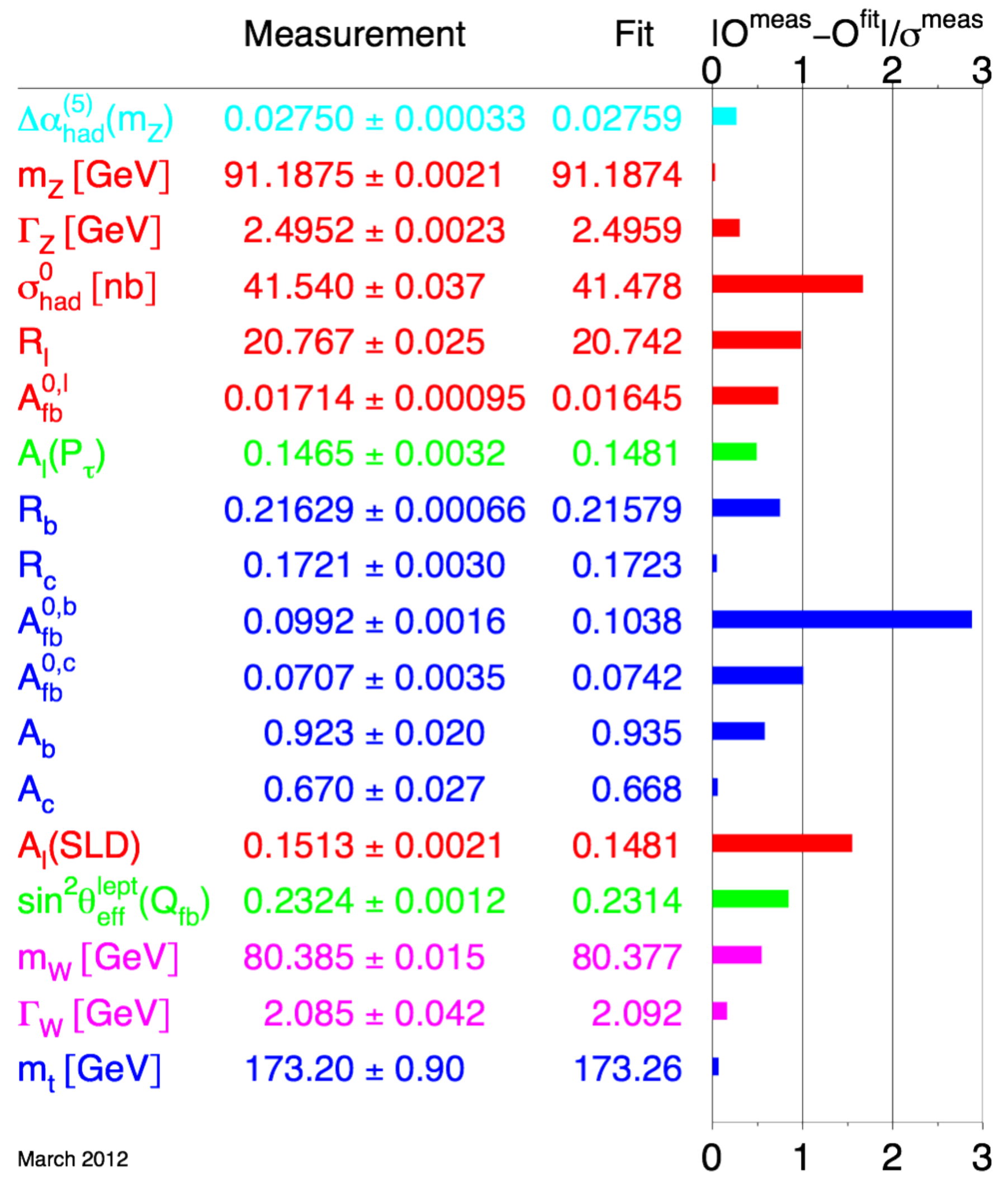}}
\caption{Comparison of measurements  
with the expectation of the SM, calculated for
the five SM input parameter values in the minimum of the global $\chi^2$ of the fit. 
The electroweak precision fit includes results from LEP (upper 13 observables), from
SLD ($A_l$) and from Tevatron ($M_W, \ \Gamma_W,\ M_t$). 
From \cite{lepewwg12} } \label{fig13}
\end{figure}

The combined precision measurements of cross sections, masses,
ratios of decay widths,  $R_\ell = \Gamma_\ell / \Gamma_{had}$,
asymmetries of angular distributions of the final state fermions from Z-boson decays (including forward-backward asymmetries like $A^{0,\ell}_{f,b}$),
and from LEP-external measurements like the left-right asymmetry from SLD,
the top-quark mass from the Tevatron proton-antiproton collider and the
mass of the W-boson from LEP phase-2 and from Tevatron,
are listed in Fig.\ref{fig13}.
They are compared to the results of a global fit of the SM predictions to
all these measurements \cite{lepewwg12}.
To the far right, the "pull"-factors, i.e. the deviation between the measurement and the
fit result, in terms of the number of standard deviations, are also shown.
The measurements agree with the SM expectations within their assigned uncertainties
and with a "normal" distribution of deviations; only three out of
18 measurements deviate by more than 1 standard deviation, and one by more than
2 (but less than 3) standard deviations.

%
%

Even though the overall picture does not indicate the need for or presence of new physics
beyond the SM,
the largest deviation, of the forward-backward asymmetry of b-quarks, $A^{0,b}_{f,b}$,
gave and still gives rise to speculations about possible reasons for it to occur.
No compelling explanation was found so far, and the origin of this deviation
is likely to be due to a statistical fluctuation or to unknown (measurement or theoretical) systematic
effects.

\begin{figure}
\centerline{\includegraphics[width=12cm]{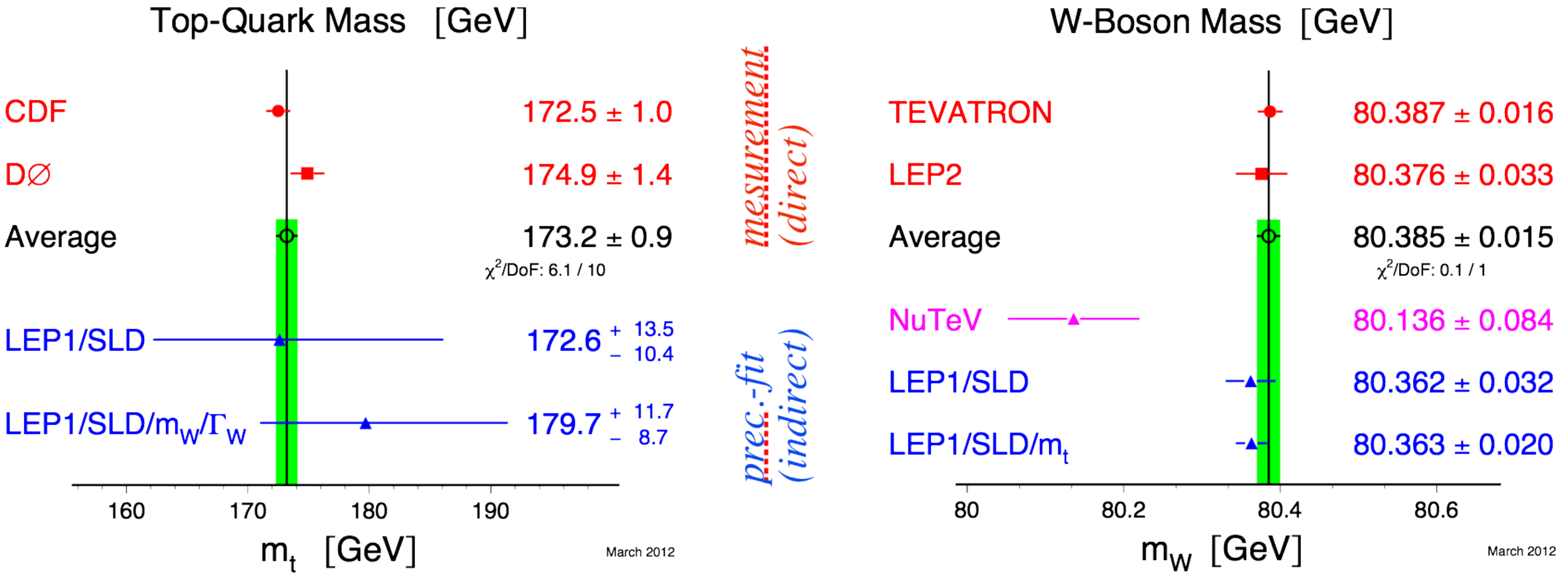}}
\caption{Results for the top-quark and the W-boson mass,
from direct measurements ($M_t$ from Tevatron experiments,
$M_W$ from Tevatron and from LEP-2) and indirect determinations
from electroweak fits to the  precision data from LEP-1 and the SLD-detector
at the Stanford Linear Collider (SLC). 
From \cite{lepewwg12} } \label{fig14}
\end{figure}

The precision of these measurements manifests in
the possibility to determine quantities like the mass of top-quark 
(discovered at the Tevatron in 1995, at the end of LEP phase-1 running), and the mass
of the Higgs-boson (discovered at the LHC in 2012, long after LEP's active phases),
from their effects through radiation corrections and quantum loop effects, see the previous 
section of this article.
This can be admired by the results shown in Fig.~\ref{fig14}, 
where the direct measurements of the masses of these objects, shown on top  and in red,
are compared with the results of the indirect results obtained from the precision fits.
While the uncertainties of the indirect fits are larger than those of the
direct measurements, the agreement between both is breathtaking - 
considering the smallness
of the radiative corrections and therefore the precision needed to obtain these results.

%
%

Similarly as for indirect determinations of the W-boson and the top-quark mass, 
significant limits for the mass of the SM Higgs boson emerged from
the electroweak precision fits, and from direct (but unsuccessful) searches for
Higgs-boson production at LEP.
The direct searches resulted in a lower limit of the Higgs-boson mass of 114.4~GeV \cite{barate03},
while
$$M_H = 129 ^{+74}_{-49}| {\rm GeV}$$
\noindent 
was obtained from the precision fits \cite{schael05} - a value that was
frequently updated and changed
within the uncertainties, but was amazingly close to the actual value
of the Higgs-boson, as known today, of $M_H = 125.09 \pm 0.24$~GeV \cite{patrignani17}.

\begin{figure}
\centerline{\includegraphics[width=11cm]{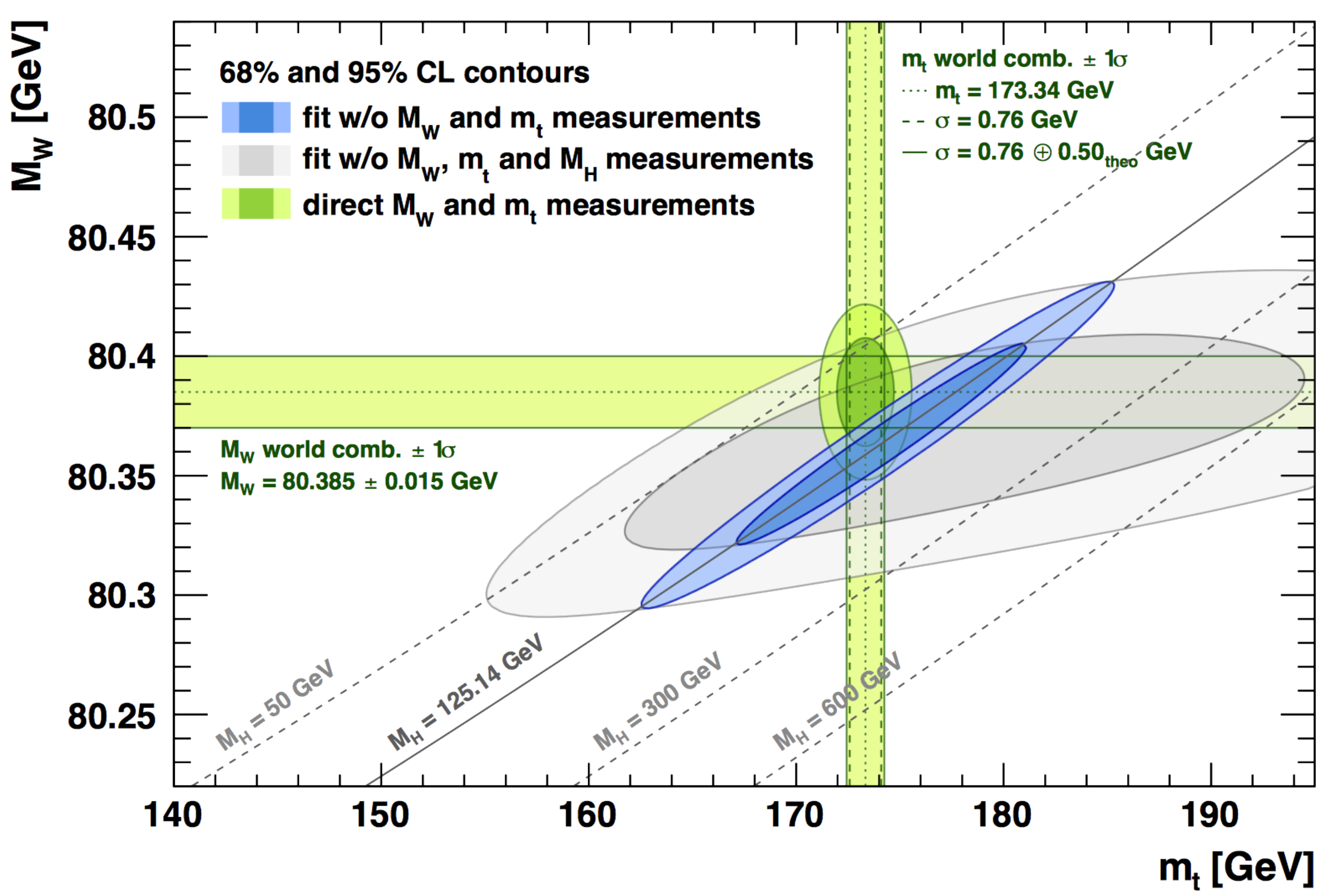}}
\caption{ 
Comparison of direct measurements of the mass of the top-quark
and of the Higgs-boson (vertical and horizontal green bands)
with indirect determinations (fits to precision data) excluding 
direct measurements (grey shaded contours) and 
including the measured Higgs-boson mass (blue shaded contours).
From \cite{gfitter14} } \label{fig15}
\end{figure}

While it was important to obtain early indications about
the possible mass range of the missing, at that time, ingredients
and objects of the SM,
these results today, with the top-quark and the Higgs-boson discovered and directly measured, still serve as important checkmarks for the consistency and reliability
of the SM.
This consistency is also obvious from Fig~\ref{fig15},
where the more recent status of indirect precision fits and direct mass measurements,
now also including the value of the Higgs-boson mass from the experiments at the
LHC, are displayed \cite{gfitter14}.
It is intriguing to see that the SM indeed still is in excellent shape!

%
%
%

%
%
\begin{figure}
\centerline{\includegraphics[width=9cm]{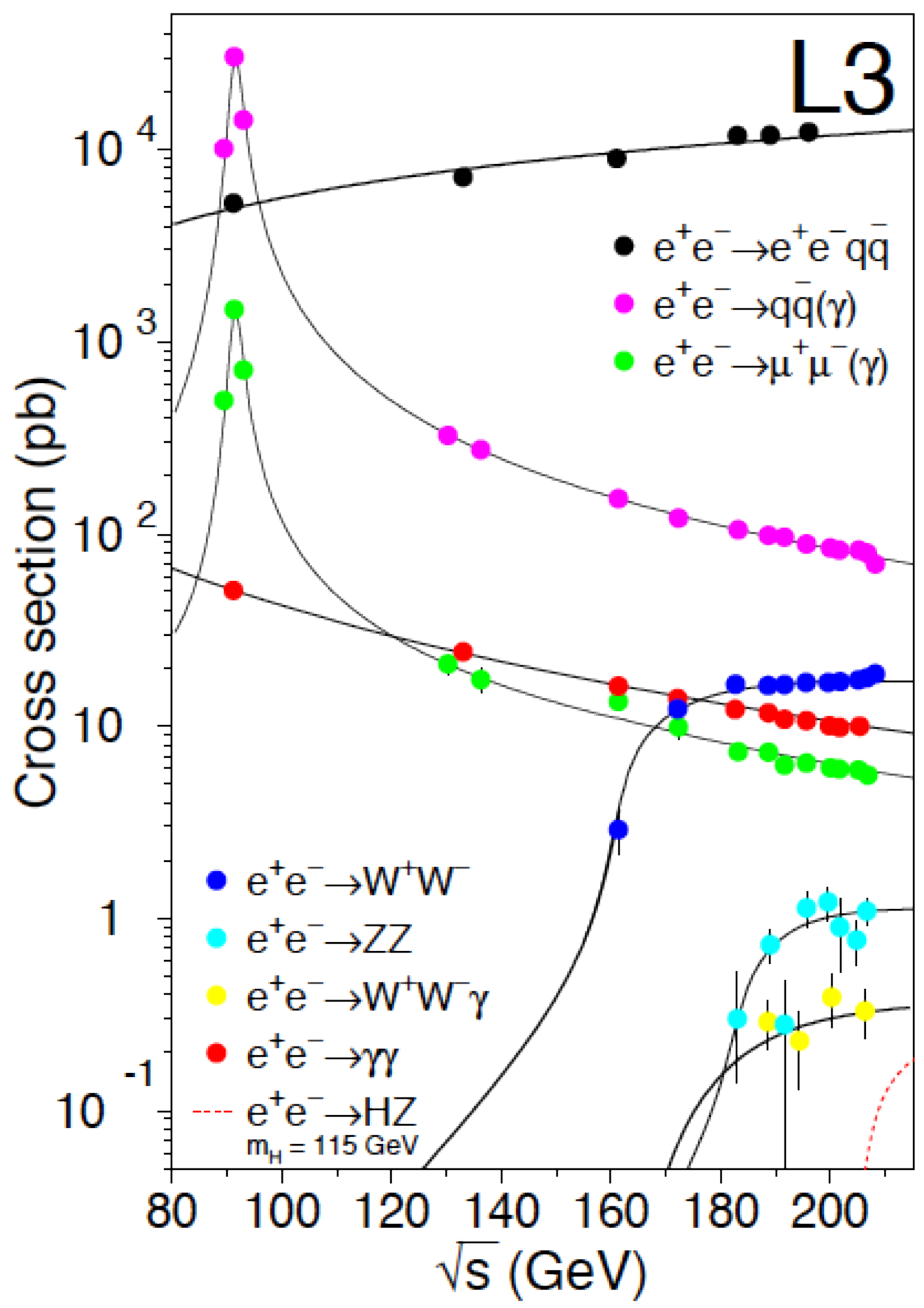}}
\caption{ 
Cross sections of electroweak SM  processes observed at LEP phase-1 and
phase-2. The curves show the theoretical predictions based on the SM.
From \cite{schael13} } \label{fig16}
\end{figure}

The success of the SM continued at LEP phase-2, at c.m. energies above the
Z-boson resonance \cite{schael13}, which already became obvious from the results including 
the LEP measurement of the W-boson mass discussed above.
The full beauty of agreement of s-dependent, measured cross sections of
virtually all processes  predicted by the SM, in the accessible range of c.m. energies,
is further demonstrated in Fig.~\ref{fig16}.
Note that the processes $e^+e^- \rightarrow e^+e^- q \overline{q}$ (c.f. Fig.~\ref{fig9}) and
$e^+e^- \rightarrow \gamma \gamma$ do not see the Z-boson resonance.
Also note that, unfortunately for those day's LEP-enthusiasts, the Higgs-boson -
now known to have a mass of $M_H = 125.09 \pm 0.24$~GeV - was just outside 
of the reach of LEP\footnote{The main production mode of Higgs bosons would have been
Higgs-Strahlung off the intermediate Z-boson, $e^+e^- \rightarrow Z H$, which requires 
c.m. energies of at least $M_Z + M_H \sim 216$\ ~GeV to obtain viable cross sections,
while the ultimate LEP energy was 209~GeV.} .
The measured shape of the W-boson pair cross section directly proved the
existence of the electroweak triple-gauge couplings (c.f. the left-most diagram of Fig.~\ref{fig7}),
and the mass and width of the W-boson was precisely measured, at LEP phase 2, to
\begin{eqnarray}
M_W &=& 80.376 \pm 0.33\ \rm{GeV,\ and} \nonumber \\
\Gamma_W &=& 2.195 \pm 0.83\ \rm{GeV}. \nonumber
\end{eqnarray}

Further measurements and combinations of results obtained at LEP phase 2 are
summarised in \cite{schael13}.

\section{Precision QCD results from LEP}

LEP data also were a source for precision studies of the strong interactions and its
underlying theory, Quantum Chromodynamics (QCD), which is regarded
to be part of the Standard Model of particle physics.
QCD also and very specifically was at the heart of Guido Altarelli.
In fact he belongs to the "fathers" of this theory with epoch-making publications like \cite{altarelli77},
and he summarised and reviewed the theoretical and experimental status of QCD at many instances,
like \cite{altarelli89b} (before the advent) and \cite{altarelli02} (after shutdown of 
LEP).

%
%
\begin{figure}
\centerline{\includegraphics[width=8cm]{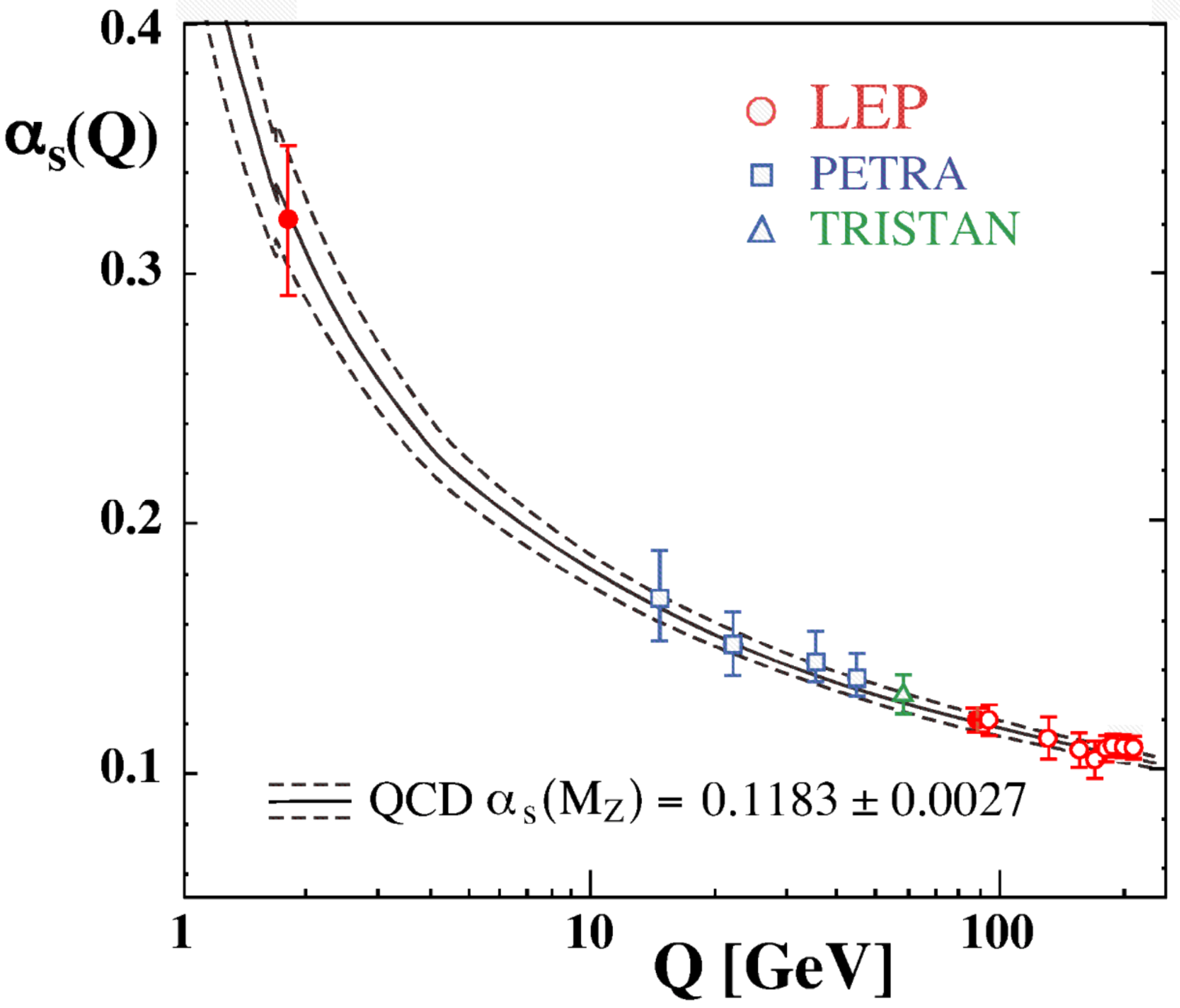}}
\caption{ 
Summary of measurements of $\alpha_s (Q \equiv E_{cm})$ from LEP. 
Results from previous $e^+e^-$-colliders are also included. 
Open symbols are from event shapes in resummed NLO, filled symbols from $\tau$ and Z-boson hadronic decay widths, in full NNLO QCD. 
The curves represent the QCD predictions of the running coupling for the world average value
of $\alpha_s(M_Z)$ at that time.
From \cite{bethke04} } \label{fig17}
\end{figure}

The spectrum of QCD studies at LEP was summarised e.g. in \cite{bethke04}.
Here, only precision determinations of the strong coupling $\alpha_s$ shall be 
briefly reviewed.
At LEP, the strong coupling was determined from a variety of observables and their
respective QCD predictions. 
The most significant determinations were obtained from measurements of hadronic event
shapes and jet rates, from the hadronic width and the spectral functions of $\tau$-lepton decays, 
and from the hadronic decay width of the Z-boson or the fits of the electroweak precision
data.
A summary of the results available in 2004 is reproduced in Fig.~\ref{fig17}.

Almost all QCD calculations and predictions that were available 
at the time of LEP running have been
significantly enhanced, from next-to-leading order in perturbation theory 
(NLO, ${\cal O}(\alpha_s^2)$) at the beginning of LEP,
to next-next-to-leading order (NNLO, ${\cal O}(\alpha_s^3)$) 
plus resummation in next-to-leading-logarithmic approximation (NLLA), 
and even to N$^3$LO in some cases, at the time thereafter until today. 
Since then, the results summarised in \cite{bethke04} have been reanalysed and 
significantly improved, see the QCD review in  \cite{patrignani17} for the most recent update
of $\alpha_s$ determinations at LEP and elsewhere.
From this, the following main results of $\alpha_s$, based on LEP data only, can be derived:
\begin{eqnarray}
{\rm from\ } \tau \ {\rm decays:\ \ } \alpha_s (M_Z) &=& 0.1192 \pm 0.0018\ {\rm (in\ N^3LO)} 
\nonumber \\
{\rm from\ electroweak\ precision\ fits:\ \ } \alpha_s (M_Z) &=& 0.1196 \pm 0.0030\ {\rm (in\ N^3LO)} 
\nonumber \\
{\rm from\ hadronic\ event\ shapes\ and\ jets:\ \ } \alpha_s (M_Z) &=& 0.1196 \pm 0.0036\ {\rm 
(in\ NNLO)} \nonumber 
\end{eqnarray}

These results are remarkably consistent with each other, with practically no correlations
between the three different classes.
They are also consistent with the current world average, $\alpha_s(M_Z) = 0.1181 \pm 0.0011$
 \cite{patrignani17}, although systematically lying on the higher side.
 This indicates that other results that enter the determination of the world average,
 like e.g. from deep inelastic scattering, but also from other studies involving a broader 
 range of (LEP and non-LEP) $e^+e^-$ data and alternative analysis methods, 
 partly arrive at smaller $\alpha_s$ values than those from LEP given above;
 see the discussion in  \cite{patrignani17}.
 
Note that the most precise results obtained at LEP, from hadronic $\tau$- and Z-boson decays,
at the smallest energy scale of $M_\tau$ and the overall reference scale of $M_Z$,
provide a very significant proof of the running of $\alpha_s$ as predicted by 
QCD, and thus, of asymptotic freedom of quarks and gluons, 
excluding a hypothetical assumption of an energy $in$dependent coupling by more than
13 standard deviations
(6 standard deviations at the time of 2004).
These results were a decisive input for granting the Physics Nobel Prize of 2004,
"for the discovery of asymptotic freedom in the theory of the strong interaction",
to David J. Gross, H. David Politzer and Frank Wilczek \cite{nobel04}.

\section{Summary}

The LEP collider provided high statistics of electron-positron collisions at 
centre-of-mass energies around the mass of the Z boson, from 1989 to 1995,
and  continued - after upgrades to using superconducting accelerator cavities - 
at energies above the Z Boson, crossing the W boson pair production threshold 
at $~160$\ GeV, up to its highest c.m. energy of 209~GeV.
LEP stopped operation in the year 2000, and was replaced by the Large Hadron Collider
being installed in the same tunnel, starting to deliver proton-proton collisions in the
(multi-)TeV c.m. energy range in 2009.

Based on the precise energy calibration and efficient operation of LEP, 
the four large general purpose detector systems ALEPH, DELPHI, L3 and OPAL,
established precision measurements of Standard Model observables and parameters 
most of which exceeded initial expectations and are unsurpassed in testing
the underlying theory and in determining its free parameters, to the level of its radiative corrections.

The LEP era manifested an overwhelming degree of glory, success and consistency 
of the Standard Model of electroweak and the strong interactions.
This review could only summarise and point out a few and the most important 
measurements and determinations.
The story of success of the Standard Model continues with the results from LHC,
demonstrating its validity up to the multi-TeV range and possibly even beyond,
perhaps reaching out up to the Planck-scale.
This success goes together with the unexpected absence of any significant signal
for new physics beyond the Standard Model, which already started to emerge 
at LEP, too.
Such a situation was considered "{\it possible but not natural}" by Guido Altarelli \cite{altarelli10}.
In the absence of direct new physics signals at current and future collider projects,
precision measurements and calculations that can reach to energy scales far beyond 
those realised in the laboratory will remain to be indispensable tools to further approach 
our ultimate understanding of  the dynamics and origin of matter and forces.

\end{document}